# Disordered magnetism in the double perovskite LaCaScIrO$_6$ with a distorted fcc lattice of Ir$^{4+}$

M. Vogl[1],* G. Bastien[1], R. Sarkar[2], R. Morrow[1], J. A. T. Barker[3], J.-C. Orain[3], H. Luetkens[3],

A.U.B. Wolter[1], H.-H. Klauss[2], S. Wurmehl[1;2], S. Aswartham[1],# and B. Büchner[1;2]

[1]Leibniz Institute for Solid State and Materials Research, D-01069 Dresden, Germany

[2]Institut für Festkörperphysik, TU Dresden, D-01062 Dresden, Germany

[3]Laboratory for Muon Spin Spectroscopy, Paul Scherrer Institute, CH-5232 Villigen PSI, Switzerland

(Dated: October 10, 2019)

**Abstract**

The synthesis and characterization of the previously unknown material LaCaScIrO$_6$ is reported. LaCaScIrO$_6$ presents a new example of the rare case of a double perovskite with the strongly spin orbit coupled 5$d$-ion Ir$^{4+}$ as its only magnetic species, forming a monoclinically distorted version of the frustrated fcc lattice. Magnetization measurements show a weak anomaly at 8 K. The Curie- Weiss temperature ϴ$_{CW}$ and effective magnetic moment $\mu_{eff}$ of LaCaScIrO$_6$ are in close proximity to the related compound La$_2$MgIrO$_6$ but differ from La$_2$ZnIrO$_6$. This suggests that the nature of the non-magnetic B-ion, namely its d-orbital filling has a strong influence on the magnetic properties. The $d^0$-ions Sc$^{3+}$ and Mg$^{2+}$ allow a different kind of exchange interactions within the Ir-sublattice than the $d^{10}$ ion Zn$^{2+}$. In addition, ac-susceptibility data does not show signs of a spin-glass ground state. The nature of the magnetism in LaCaScIrO$_6$ has been further elucidated using muon spin relaxation measurements. The zero-field measurements reveal the absence of well-defined oscillations down to 1.6 K, while temperature dependent µSR studies show an anomaly at 8 K. Overall, our result suggest the presence of two different magnetic environments or domains in LaCaScIrO$_6$, which is likely related to its structural features.

*m.vogl@ifw-dresden.de
#s.aswartham@ifw-dresden.de

## I. INTRODUCTION

Transition metal oxides of the 5d elements have been subject of intense research over the last two decades. Their unique interplay of crystal field splitting, large spin-orbit coupling (SOC) and Coulomb repulsion has lead to intriguing new ground states [1-3]. One of the main driving forces for the experimental interest has been theoretical predictions of novel magnetically frustrated phases like quantum spin liquids and spin ice states. In particular, iridates with $Ir^{4+}$-states, which possess an intriguing $j_{eff}$ = 1/2 state caused by the aforementioned interactions, have been in the focus of experiments, attempting to realize the theoretical predictions. In the field of geometrically frustrated systems, the honeycomb lattice, as the ideal realization of the Kitaev Hamiltonian, has been in the focus of studies on iridates like $Li_2IrO_3$ and $Na_2IrO_3$ as well as $\alpha$-$RuCl_3$ [4-7]. Another geometrically frustrated lattice that has been found to host Kitaev interactions is the face-centered cubic (fcc) lattice [8-11]. Magnetic fcc sublattices can be realized in double perovskite structure type materials of the general formula $A_2BB'O_6$ [11; 12]. They consist of corner-sharing $BO_6$- and $B'O_6$-octahedra with the A-cations sitting in the void space in between the octahedra. In the most common "rock-salt"-substructure type, B and B'-cations are alternating. In "rock-salt"-ordered double perovskites the B- and B'-sublattices can be seen as two interpenetrating fcc-lattices. If only one B-cation owns a magnetic moment, while the 2nd B-position, as well as the A-sites, are occupied by non-magnetic cations, the magnetic properties of the compound should be solely determined by the magnetic cations, which are located on the fcc sublattice.

Known examples, in which this has been realized for $Ir^{4+}$ as the sole magnetic ion, include $La_2ZnIrO_6$, $La_2MgIrO_6$, $Sr_2CeIrO_6$ and $Ba_2CeIrO_6$. $Ba_2CeIrO_6$ has a cubic unit cell, but octahedral distortions lead to local structural differences which release the magnetic frustration and lead to a long-range order [10-13]. In the cases of $Sr_2CeIrO_6$, $La_2ZnIrO_6$ and $La_2MgIrO_6$, no perfect cubic perovskite structure is present. Instead, a monoclinic distortion and octahedral tiltings lead to a deviation of the fcc-sublattice of $Ir^{4+}$ [14-17], so that the $Ir^{4+}$- sublattice can be viewed as a monoclinically distorted fcc lattice. Despite possessing very similar crystal structures, $La_2ZnIrO_6$, $La_2MgIrO_6$ and $Sr_2CeIrO_6$ have been found to adopt different magnetic ground states. $La_2ZnIrO_6$ is a canted antiferromagnet ($T_N$ = 7 K) while $La_2MgIrO_6$ and $Sr_2CeIrO_6$ show collinear antiferromagnetism at 12 K and 21 K, respectively. As the minor structural differences, specifically within $La_2(Zn,Mg)IrO_6$, do not offer an explanation for the discrepancy of the magnetic properties, the reason assumingly lies in the nature of the non-magnetic B-cation itself. This phenomenon has been studied in detail for osmates, such as $Sr_2BOsO_6$ (B = Y, In, Sc), in which a correlation of the transition temperatures of the materials and the electronic configuration of the non-magnetic B-ion (d0 vs. d10) was found [18]. For double perovskite iridates, similar effects have been suggested as an explanation for the different magnetic properties of $La_2ZnIrO_6$ and $La_2MgIrO_6$ [11]. However, additional examples are desirable in order to establish empirical trends in fcc-iridates.

In this work we report on the synthesis and magnetic properties of the previously unknown material $LaCaScIrO_6$, a new monoclinic double perovskite with an $Ir^{4+}$-based quasi-fcc lattice as it's only magnetic species and a disordered A-site, shared by $La^{3+}$ and $Ca^{2+}$. We compare our findings with the known double perovskites $La_2ZnIrO_6$ and $La_2MgIrO_6$.

## II. EXPERIMENTAL DETAILS

Polycrystalline $LaCaScIrO_6$ was synthesized from stoichiometric amounts of $La_2O_3$, $CaCO_3$, $Sc_2O_3$ and $IrO_2$. $La_2O_3$ was heated prior to the reaction to $900^0C$ for 12 h to remove moisture. All starting materials were ground and pelletized. In a first annealing step the mixture was heated to $650^0C$ for 12 h with a heating rate of $200^0C/h$ and subsequently to $1200^0C$ with a rate of $200^0C/h$ and a dwelling time of 60 h. After one regrinding step, another annealing step followed. The mixture was reheated to $1200^0C$ with a rate of $200^0C/h$ and allowed to dwell for 72 h. The cooling was done in a similar way as for the first annealing step. For structural analysis, X-ray diffraction measurements (XRD) were carried out in the transmission method on a StoeStadi-Power diffractometer with Co- $K_{\alpha 1}$ radiation. Rietveld-refinement was performed using the FullProf software package [19; 20]. Scanning electron microscopy (SEM) and energy dispersive x-ray spectroscopy (EDXS) were performed on a nanoSEM by FEI. The analysis was carried out on pressed powder pellets.

Magnetic measurements were performed using a Quantum Design MPMS-XL SQUID-magnetometer. The ac magnetic susceptibility was measured on a powder sample of about 200 mg. The specific heat of $LaCaScIrO_6$ was measured on a pressed pellet with a commercial PPMS calorimeter down to 1.8 K and under magnetic fields up to 9 T. The contribution of the phonons to the specific heat was estimated from the measurement of the specific heat of the non-magnetic analog $La_2ZnPtO_6$. Polycrystalline $La_2ZnPtO_6$ was synthesized from $La_2O_3$, ZnO and $PtO_2$, using the same procedure as described for $La_2ZnIrO_6$ in Ref. [21]. The double perovskite structure of $La_2ZnPtO_6$ was checked by x-ray diffraction and its diamagnetic nature was confirmed by magnetization measurements (see supp. material). Zero-field (ZF), longitudinal-field (LF), and transverse-field (TF) muon spin relaxation (μSR)measurements were performed on the General Purpose Spectrometer (GPS) and the Dolly instrument at the Swiss Muon Source (SμS), Villigen-PSI, Switzerland. A powder sample was pressed into a pellet (8mm in diameter), enclosed in pouch constructed from Mylar tape, and mounted on a copper fork designed to maximize muons stopping in the sample. On Dolly, the sample was placed in an Oxford Instruments Varioxow cryostat with a base temperature of ≈ 1.6 K, and on GPS a QUANTUM flow cryostat with a similar base temperature was used.

III. RESULTS

**A. X-ray diffraction and SEM/EDXS-analysis**

The crystal structure of $LaCaScIrO_6$ was analyzed by x-ray powder diffraction. The resulting pattern is shown in Fig. 1. Rietveld refinement confirms that the main phase adopts a monoclinic double perovskite structure with the space group *P 21/n*. The octahedral tilting's associated with this space group follow the Glazer notation $a^-a^-c^+$ [22]. Besides the main phase, some additional reflections and shoulders of the main peaks are observed. These can be attributed to a foreign phase, which has a crystal structure in the space group P n m a. This space group is typical for single perovskites following the tilt system $a^-a^-c^+$, the same one as found for the main phase double perovskite. The single perovskite most likely to be present as an impurity phase in this case is $LaScO_3$. Its portion is estimated to be approximately 6% of the sample by Rietveld refinement. Similar to $La_2ZnIrO_6$, $La_2MgIrO_6$ and $Sr_2CeIrO_6$, no cubic double perovskite structure is obtained for $LaCaScIrO_6$. Hence, the magnetic Ir-sublattice forms a monoclinically distorted fcc lattice (Fig. 2).

To probe the anti-site disorder between the B-sublattices, the occupations of Sc and Ir were refined. The best fit was obtained for an anti-site disorder of 9%. B-site disorder is a common feature in double perovskites. The degree of disorder highly depends on the ionic radii and the charge ratio of the B-cations. In the case of $Sc^{3+}$ (r = 0.745 Å) and Ir4+ (r = 0.625 Å), both are relatively close, so a disorder of 9% is not surprising. La and Ca are occupying the same crystallographic site. Therefore an estimation of the La/Ca-disorder is not possible via Rietveld refinement. The structure model indicates a random distribution of both A-cations. A strongly ordered arrangement would lead to additional supercell reflections, which are not found in the XRD data. In general, A-site order in AA'BB'O6 perovskites is difficult to achieve. Only a few examples for fully A- site ordered perovskites are known. For instance, one common strategy to achieve full A-site order requires an alkaline earth metal as one A-cation and a large difference in the oxidation state between the B-cations, often including $W^{6+}$ on one of the B-sites [23-25].

In $LaCaScIrO_6$ the conditions commonly associated with A- site order are not met, therefore a random distribution of $La^{3+}$ and $Ca^{2+}$ is expected. The chemical composition of the sample has been analyzed by SEM and EDXS. In Fig. 3, an SEM image taken with a secondary electron (SE) detector showing the sample topography is shown alongside an image taken with a backscattered electron (BSE) detector unveiling chemical contrast. The SE image shows a rough sample surface. Achieving a smoother surface is difficult for double perovskite iridates since the pressed powder pellets used in the SEM analysis are brittle and cannot be thoroughly polished. In the BSE image, no significant color differences that cannot be traced back to the uneven surface are found, which indicates good chemical homogeneity. The composition was measured at several points of the sample to achieve reasonable statistics. According to this data, the composition of the sample is $La_{1.04}Ca_{1.03}Sc_{1.02}Ir_{0.91}O_6$, which is within the range of the nominal composition. The oxygen content has not been determined in this analysis due to its high inaccuracy when measured by EDXS.

**B. Magnetic properties**

The temperature dependent magnetization of $LaCaScIrO_6$ is shown in Fig. 4. A weak anomaly is observed at $T_N \approx 8$ K. Additionally there is a splitting between field-cooled (FC) and zero-field cooled (ZFC) curves at 10 K. This leads to the anomaly being more pronounced in the ZFC data. A Curie-Weiss analysis in the temperature range 100 K $\leq T \leq$ 280 K yields a Curie-Weiss temperature of $\theta_{CW}$ = -22.1 K and an effective magnetic moment $\mu_{eff}$ = 1.51 $\mu_B$ (Fig. 5). The value for $\mu_{eff}$ is close to the ones found for $La_2ZnIrO_6$ (1.71 $\mu_B$) and $La_2MgIrO_6$ (1.42 $\mu_B$) and in proximity to the theoretical value for the $j_{eff}$ = ½ state of $Ir^{4+}$ (1.73 $\mu_B$) [16]. The negative Curie-Weiss temperature indicates dominant antiferromagnetic interactions. In order to investigate the possible occurrence of a spin glass state at low temperature, we performed ac magnetic susceptibility measurements. The ac magnetic susceptibility as a function of temperature and frequency is represented in Fig. 6. Static and alternating fields of 1000 Oe and 5 Oe were respectively applied. The ac magnetic susceptibility is independent of the frequency within the experimental error bars and undergoes a temperature dependence very similar to the one of the magnetization reported in Fig. 4. On the contrary a maximum of the magnetic susceptibility at a frequency dependent freezing temperature would be expected for a spin glass state. Thus this ac magnetic susceptibility measurement rules out the possibility of a spin glass state at low temperature in $LaCaScIrO_6$.

**C. Specific heat**

The specific heat of $LaCaScIrO_6$ is the sum of a phononic and a magnetic contribution (Fig. 7). The phononic contribution to the specific heat of $LaCaScIrO_6$ is rather diffcult to estimate considering the random distribution of two atoms with different masses, La and Ca, on the same crystallographic site. As a simple approach, the phononic contribution to the specific heat in LaCaScIrO$_6$ was obtained from scaling the specific heat of non-magnetic La$_2$ZnPtO$_6$, neglecting the effect of disorder. The scaling factor (Lindemann factor) was computed according to Refs. [26; 27]. Taking into account both the difference in molar mass and molecular volume between the two compounds, we obtained the Lindemann scaling factor: $\Theta_{D(LaCaScIrO6)}$ / $\Theta_{D(La2ZnPtO6)}$ = 1.12. The resulting magnetic contribution to the specific heat of LaCaScIrO$_6$ divided by the temperature is represented in Fig. 7(b). No sharp magnetic transition is observed in the specific heat down to 2 K. Instead, a broad maximum of C$_p$mag/T as a function of temperature is discernible around 8 K in line with the dc and ac susceptibility studies. The application of a magnetic field slightly shifts this maximum to 7.2 K at 9 T, showing that no long-range order can be induced in LaCaScIrO$_6$ by the application of an external magnetic field up to 9 T. The magnetic entropy change $\Delta S_{mag}$ of LaCaScIrO$_6$ in absence of a magnetic field as a function of temperature is represented in Fig. 7(c). The total magnetic entropy change S$_{mag}$(25 K) - S$_{mag}$(2 K) $\simeq$ 2 J/mol/K is much smaller than the value corresponding to the ordering of j$_{eff}$ = 1/2 magnetic moments $Rln$(2) = 5.8 J/mol/K, suggesting some residual magnetic entropy at low temperatures T ≤ 2 K and/or an overestimated entropy value due to the potential presence of non-magnetic Ir$^{3+}$- and Ir$^{5+}$-ions in the sample (μSR part, see below).

**D. muon spin resonance**

To understand the system from the microscopic point of view, we have performed detailed μSR investigations on this system. μSR is proven to be an ideal tool to understand the local magnetism. Zero-field (ZF), longitudinal-field (LF), and transverse-field (TF) muon spin relaxation (μSR) measurements are performed and presented here. Measurements performed in a weak transverse-field can be used to determine the magnetic volume fraction in a given sample. Muons stopping in a magnetic phase are rapidly depolarized, and do not contribute to the precession signal at long times. Muons stopping in a paramagnetic region will precess at approximately the applied frequency, with a relaxation rate determined by the local electronic and nuclear-dipolar environment. The ratio of the amplitude of the oscillating signal(A$_{osc}$) and the full asymmetry measured well outside the magnetic phase (A$_0$) gives the paramagnetic/nonmagnetic volume fraction, f$_{PM}$ = A$_{osc}$/A$_0$. The magnetic volume fraction is then given by f$_M$ =1-f$_{PM}$. μSR time-spectra were collected in a transverse-field of 3mT in the temperature range 3K ≤ T ≤ 15 K and f$_M$ was calculated for each temperature. Two time-spectra collected at 3K and 14K are presented in Fig. 8(a). The signal oscillates with its full amplitude at 14 K. Reducing the temperature to 3K leads to a reduction of the precession signal by a factor of around 2/3. The full temperature dependence of f$_M$ is presented in Fig. 8(b). The volume fraction gradually decays to 0 for temperatures >15 K, with the midpoint of the transition at ≈ 8.41 K, which is very close to the separation between the ZFC and FC magnetic susceptibilities. The maximum value of the volume fraction extracted from the logistic fit is ≈ 65% indicating that around 1/3 of the sample remains paramagnetic in a field of 3 mT.

To make an estimation about the number of possible muon sites present in this compound further transverse-field measurements (0.78 T) have been carried out. Experimentally, one can attempt to distinguish the number of muon sites by measuring the Knight shifts in a given material. The Knight shift depends on the local dipolar-field coupling tensor, and different unique muon sites would be expected to produce different Knight shifts. Therefore, the presence of multiple peaks in the Fourier transform of the high-field time-spectra may suggest how many unique muon sites are in a given sample. The resolution of this procedure is limited in powder samples, as the line-shapes are significantly broadened due to polycrystalline averaging. Fig. 8(b) (inset) shows the resulting field distribution at two different temperatures revealing a single peak at each temperature. Both spectra are well described by a single exponentially-damped cosine oscillation. This suggests that most likely only one muon site exists in LaCaScIrO$_6$. Fig. 9 presents the observed time-spectra of the muon spin polarization function $G_Z^{ZF}$ (t)after zero-field cooling for T =1.6, 9.0 and 40 K. At all three temperatures, there is no evidence of any well-defined precession frequency suggesting the absence of well-defined static internal field at the muon site which may be expected for a system exhibiting long-range order. This observation suggests that if a static magnetic phase is present, then it is spatially disordered. The two-component structure (a fast and a slow relaxing one) evident in Fig. 9 persists for T ≤ 15 K. The best model describing the data was given by

$$G_Z^{ZF}(t) = f_M^{ZF} G_Z^{KT}(t, \sigma_{KT})e^{-\Lambda_f t} + (1 - f_M^{ZF})e^{-(\Lambda_s t)^\beta}. \quad (1)$$

Here, $f_M^{ZF}$ and (1- $f_M^{ZF}$) represents the fraction of the fast and slowly relaxing components of the signal, respectively, with $f_M^{ZF}$ constrained to lie between 0 and 1 during the fitting routine, and $G_Z^{KT}$ is the static Gaussian Kubo-Toyabe function, describing randomly oriented moments with a Gaussian field-width σ$_{KT}$. The exponential relaxation multiplying the $G_Z^{KT}$ indicates that the 1/3 recovery expected for a static distribution is significantly damped by fluctuations of the local moment.

Fig. 10 (a, b) displays the temperature dependence of the Gaussian Kubo-Toyabe relaxation rate σ$_{KT}$ and the fast-decaying rate Λ$_f$ . The fact that a Gaussian distribution is observed at low temperatures indicates that the magnetism is disordered. As the temperature is increased above 7.5 K, there is a crossover from Gaussian to exponential relaxation, signaling the loss of the static spin component, and the onset of strong thermal fluctuations in the local field of the 'frozen spin' domains. The so called decoupling experiments, that is, measurements made in a magnetic field applied parallel to the initial muon spin polarization direction are presented in Fig. 11. The spectra in the figure are fitted to a simple model

$$G_Z^{LF}(t) = f_G \exp(-1/2\sigma^2 t^2) + (1 - f_G) \exp(-\Lambda_{LF} t). \quad (2)$$

Time-spectra in the time range 0-0.5 µs clearly demonstrate that the Gaussian part of the time-spectra is completely decoupled in fields larger than 2500 G. This further hints at the static nature of the internal field distribution. Fig. 10 (c) displays the slow-component exponential relaxation rate. As T → 8K from high temperatures, there is an increase $\Lambda_s$. In the high temperature range the local moments (spins) are fluctuating very fast. With lowering the temperature a critical slowing down effects is relevant. Thus, $\Lambda_s$ increase with lowering temperatures.

The existence of two different relaxation components in zero-field data suggests two different magnetic environments or domains. The fast-relaxing Gaussian component existing for T ≤ 15K most likely arises due to `frozen spin' domains, in which the static or quasi static internal field distribution has zero-mean and a second moment $\Delta B = \sigma_{KT} / \gamma_\mu$, where $\gamma_\mu$ is the muon gyromagnetic ratio. For a randomly oriented distribution of moments, the zero-field µSR spectrum should be described by the static Gaussian Kubo-Toyabe function $G_{KT}(t)$. We have success in applying this model to the data, but only upon multiplying by a further exponential function with decay rate $\Lambda_f$. The effect of this decay is primarily in suppressing the `1/3 tail' of the static Kubo-Toyabe function, and indicates that the recovery signal is damped by spin-lattice relaxation. The low-temperature value of $\sigma_{KT}$ observed here corresponds to an internal field width $\Delta B = (19.6 \pm 0.4)$ mT.

The XRD data for this sample show the expected distorted perovskite structure with a monoclinic unit cell alongside a small impurity. Given the significant volume fraction that has been measured, the paramagnetic domains are unlikely to arise from the impurity. The disordered magnetism that has been observed could arise due to lattice defects, due to the half-occupancy of $La^{3+}$ and $Ca^{2+}$ ions at the 4e Wyckoff position.

IV. DISCUSSION

Our study of $LaCaScIrO_6$ yields interesting insight into the exchange interaction in double perovskites with only a single magnetic B-ion. A comparison to previously known iridates of this kind, specifically $La_2ZnIrO_6$ and $La_2MgIrO_6$, shows that all three possess a negative Curie-Weiss temperature, hence antiferromagnetic exchange is dominant in all three materials. The absolute value of $\theta_{CW}$ for LaCaScIrO$_6$ (-22 K) however is much closer to the one of La$_2$MgIrO$_6$ (-24 K) than La$_2$ZnIrO$_6$ (-3 K) [15; 16]. Since $\theta_{CW}$ can be seen as a measure of the interaction strength between the magnetic atoms in a material, in this case between the Ir$^{4+}$ -ions, it allows to draw conclusions about the underlying exchange mechanism. In the double perovskite osmates Sr$_2$BOsO$_6$ (B = Y, In, Sc), that have 5d$^3$ Os$^{5+}$ as their only magnetic ion, it was found that the d-orbital configuration of the non-magnetic B-ion strongly influences the exchange within the magnetic B-sublattice, i.e. for B = Sc/Y, which are $d^0$-ions, a significantly higher ordering temperature is found than for the $d^{10}$-ion In$^{3+}$[18]. The d-orbital configuration of the non-magnetic B-ion also offers a possible explanation for the stronger similarity between LaCaScIrO$_6$ and La$_2$MgIrO$_6$ as compared to La$_2$ZnIrO$_6$. Sc$^{3+}$ and Mg$^{2+}$ have a $d^0$-configuration while Zn$^{2+}$ is a $d^{10}$-ion. The underlying reason behind this effect may lie in the orbital hybridzation. The $d^0$- ions allow strong hybridization of their d-orbitals with the neighboring oxygen 2p-orbitals. This allows a superexchange pathway going through the $d^0$-ions. On the other hand a filled $d^{10}$-shell as in Zn$^{2+}$ prevents hybridization and disables this exchange pathway. Similar effects have been proposed as the cause of a high degree of frustration in Sr$_2$Cu(Te$_{0.5}$W$_{0.5}$)O$_6$ [28].

The striking difference between LaCaScIrO$_6$ and its brother compounds La$_2$MgIrO$_6$ and La$_2$ZnIrO$_6$ lies in the lack of long-range magnetic order in the system, which is indicated by the absence of a sharp anomaly in the specific heat data. Consequently, the anomaly found in the magnetization at 8 K may correspond to the onset of short-range magnetic phenomena. The presence of a spin-glass state however can be excluded as the typical frequency dependency associated with spin-glasses is not found in the ac-susceptibility. The μSR investigations on LaCaScIrO$_6$ found disordered static magnetism below a temperature of ≈ 8 K, in agreement with the anomaly temperature in the magnetization data. In addition, the two-component relaxation observed in μSR suggests an inhomogeneous distribution of Ir$^{4+}$ magnetism over two types of magnetic domains. In many systems, the existence of short-range correlations that fail to stabilize a long-range order is accompanied by disorder in the atomic structure [29-32]. In the case of LaCaScIrO$_6$, this disorder can likely be found on the crystallographic A-site. A random distribution between La$^{3+}$ and Ca$^{2+}$, which is likely according to x-ray diffraction studies, can induce local differences in the atomic structure, affecting octahedral tiltings and bond angles within the B-sublattices, which in turn influence the magnetic exchange between neighboring Ir$^{4+}$-ions. This may lead to local differences of the magnetic exchange pathways which may cause spatially limited, short-range ordered domains, but prevents full long-range order, causing the system to adapt an exotic ground state. As an additional effect, the disorder between La$^{3+}$ and Ca$^{2+}$ can lead to a local accumulation of either of the two A-cations, affecting the oxidation state of nearby Ir-ions. A higher amount of La$^{3+}$/Ca$^{2+}$ in its vicinity would cause iridium to adopt the oxidation state $Ir^{3+}/Ir^{5+}$ instead of the nominal $Ir^{4+}$. Both $Ir^{3+}$ and $Ir^{5+}$ are non-magnetic. Therefore their presence would dilute the $Ir^{4+}$-sublattice, increasing the distances between the magnetic ions and counteract long-range magnetic order.

## V. CONCLUSIONS

Polycrystalline double perovskite $LaCaScIrO_6$ has been successfully synthesized. It presents a previously unknown example of $Ir^{4+}$ on a geometrically frustrated quasi-fcc sublattice. Magnetization data shows a small anomaly at $T_N \approx 8$ K. Comparison of our findings with the similar compounds $La_2ZnIrO_6$ and $La_2MgIrO_6$ suggest a closer affinity with $La_2MgIrO_6$. This result agrees with a theoretical work on osmates which suggests an influence of the *d*-orbital configuration of the non-magnetic B-cation on the magnetic properties of quasi-fcc double perovskites. Specific heat and µSR studies reveal the lack of a long-range magnetic order in $LaCaScIrO_6$. A spin-glass ground state can be excluded due to the lack of frequency dependency of the ac-susceptibility. Furthermore, µSR measurements reveal two magnetically distinct sites in the sample, likely caused by domain formation associated with lattice disorder and local structural changes. For a deeper understanding of the magnetic ground state, neutron powder diffraction studies on $LaCaScIrO_6$ are planned.

## VI. ACKNOWLEDGEMENTS

We gratefully acknowledge financial support by the Deutsche Forschungsgemeinschaft (DFG) through projects B01 and C02 (RS, HHK) of the SFB 1143 (project-id 247310070), grant AS 523/4-1 (SA) and grant WU 595/3-3; (SW). RM acknowledges support from the Alexander von Humboldt Foundation. GB acknowledges financial support from the European Union's Horizon 2020 research and innovation program under the Marie Skłodowska-Curie grant agreement No 796048. Technical assistance was provided by C. G. F. Blum, S. Gaß and S. Müller-Litvanyi. We thank A. A. Aczel for fruitful discussions.

**References**

1. B. J. Kim, H. Jin, S. J. Moon, J.-Y. Kim, B.-G. Park, C. S. Leem, J. Yum T.W. Noh, C. Kim, S.-J. Oh, J.-H. Park, V. Durairaj, G. Cao, E. Rotenberg, Phys. Rev. Lett., 101, 076402 (2008).

2. D. Pesin, L. Balents, Nat. Phy., 6, 376 (2010).

3. W. Witczak-Krempa, G. Chen, Y. B. Kim, L. Balents, Annu. Rev. Condens. Matter Phys., 5, 57-82 (2014).

4. A. Kitaev, Annals of Physics, 321, 2-111 (2006).

5. Y. Singh, S. Manni, J. Reuther, T. Berlijn, R. Thomale, W. Ku, S.Trebst, P. Gegenwart, Phys. Rev. Lett., 108, 127203 (2012).

6. J. Chaloupka, G. Jackeli, G. Khaliullin Phys. Rev. Lett., 105, 027204 (2010).

7. K. W. Plumb, J. P. Clancy, L. J. Sandilands, V. Vijay Shankar, Y. F. Hu, K. S. Burch, H.-Y. Kee, Y.-J. Kim, Phys. Rev. B, 90, 041112(R) (2014).

8. I. Kimchi, A. Vishwanath, Phys. Rev. B 89, 014414 (2014).

9. H. T. Diep, H. Kawamura, Phys. Rev. B 40, 7019 (1989).

10. A. A. Aczel, J. P.Clancy, Q. Chen, H. D. Zhou, D. Reig-i-Plessis, G. J. MAcDougall, J. P. C. Ru_, M. H. Upton, Z. Islam, T. J. Williams, S. Calder, J.-Q. Yan. arXiv:1901.08146 (2019).

11. A. A. Aczel, A. M. Cook, T. J. Williams, S. Calder, A. D. Christianson, G.-X. Cao, D. Mandrus, Y. B. Kim, A. Paramekanti, Phys. Rev. B 93, 214426 (2016).

12. E. Kermarrec, C. A. Marjerrison, C. M. Thompson, D.D. Maharaj, K. Levin, S. Kroeker, G. E. Granroth, R. Flacau, Z. Yamani, J. E. Greedan, B. D. Gaulin, Phys. Rev. B, 91, 075133 (2015).

13. A. Revelli, C. C. Loo, D. Kiese, P. Becker, T. Fröhlich, T. Lorenz, M. Moretti Sala, G. Monaco, F, L. Buessen, J. Attig, M. Hermanns, S. V. Streltsov, D. I. Khomskii, J. van den Brink, M. Braden, P. H. M. van Loosdrecht, S. Trebst, A. Paramekanti, M. Grüninger, arXiv:1901.06215 (2019).

14. D. Harada, M. Wakeshima, Y.Hinatsu J. Solid State Chem., 145, 356-360 (1999).

15. A.V. Powell, J. G. Gore, P. D. Battle, J. of Alloys and Comp., 201, 73 (1993).

16. G. Cao, A. Subedi, S. Calder, J.-Q. Yan, J. Yi, Z. Gai, L. Poudel, D. J. Singh, M. D. Lumsden, A. D. Christianson,
B. C. Sales, D. Mandrus, Phys. Rev. B 87, 155136 (2013).

17. A. M. Cook, S. Matern, C. Hickey, A. A. Aczel, A. Paramekanti, Phys. Rev. B 92, 020417(R) (2015).

18. S. Kanungo, B. Yan, C. Felser, M. Jansen, Phys. Rev. B, 93, 161116 (2016).

19. H. M. Rietveld, J. Appl. Cryst. 2, 65 (1969).

20. T. Roisnel, J. Rodriguez-Carvajal, Mater. Sci. Forum 118, 378 (2001).

21. M. Vogl, L. T. Corredor, T. Dey, R. Morrow, F. Scaravaggi, A. U. B. Wolter, S. Aswartham, S. Wurmehl, B. Büchner, Phys. Rev. B, 97, 035155 (2018).

22. A. M. Glazer Acta Cryst. B 28, 3384-3392 (1972).

23. M. C. Knapp, P. M. Woodward, J. Solid State Chem. 179, 1076-1085 (2006).

24. G. King, S. Thimmaiah, A. Dwivedi, P. M. Woodward, Chem. Mater. 19, 6451 6458 (2007).

25. G. King, P. M.Woodward, J. Mater. Chem. 20, 5785 5796 (2010).

26. A. Tari, The Specific Heat of Matter at Low Temperatures, Imperial College Press (2003). J. Mater. Chem. 20, 5785 5796 (2010).

27. J. W. Kim, Y. S. Oh, K. S. Suh, Y. D. Park, K. H. Kim, Thermochimica Acta 455, 2 (2007).

28. O. Mustonen, S. Vasala, E. Sadrollahi, K. P. Schmidt, C. Baines, H. C. Walker, I. Terasaki, F. J. Litterst, E. Baggio-Saitovitch, M. Karpinnen, Nature Comm. 9, 1085 (2018).

29. F. Jimenez-Villacorta, I. Puente-Orench, J. Rodriguez- Carvajal, L. H. Lewis, Materials and Design 112, 124 130 (2016).

30. J.-U. Hoffmann, D. Hohlwein, R. Schneider, A. H. Moudden, Physica B 276-278, 608 609 (2010).

31. M. Abassi, Z. Mohamed, J. Dhahri, E. K. Hlil, J. Alloy Compd 664, 657 663 (2016).

32. D. Li, Y. Homma, F. Honda, T. Yamamura, D. Aoki, Physics Procedia 75, 703 710 (2015).

**Captions**

**FIG. 1:** XRD-pattern of LaCaScIrO$_6$. The Rietveld refinement confirms the formation of a monoclinic double perovskite structure. A single perovskite phase is found as impurity.

**FIG. 2**: Crystal structure of LaCaScIrO$_6$.

**FIG. 3**: SEM-pictures of LaCaScIrO$_6$. A comparison of topography (left) and chemical contrast (right) images suggests good chemical homogeneity; for details see text

**FIG. 4**: Field-cooled (FC; closed symbols)) and zero-field cooled (ZFC; open symbols) temperature dependent magnetization of LaCaScIrO$_6$ in an external field of 10 kOe. An anomaly is observed at 8 K.

**FIG. 5**: Curie-Weiss analysis of the high-temperature magnetization of LaCaScIrO$_6$. The effective magnetic moment is in close proximity to similar Ir4+-based perovskites and to the theoretical value. The negative Curie-Weiss temperature indicates dominant antiferromagnetic interactions.

**FIG. 6**: ac magnetic susceptibility of LaCaScIrO$_6$ as a function of temperature for different frequencies under a static magnetic field of 1000 Oe and an alternating magnetic field of 5 Oe. No frequency dependence is observed within the experimental resolution.

**FIG. 7**: (a) Specific heat $C_p$ of LaCaScIrO$_6$ and of the nonmagnetic analog La$_2$ZnPtO$_6$ as a function of temperature. The estimation of the phonon contribution in LaCaScIrO$_6$ computed from the specific heat

of La$_2$ZnPtO$_6$ by applying the Lindemann scaling is plotted together. (b) Magnetic contribution to the specific heat of LaCaScIrO$_6$ divided by the temperature $C_{pmag}/T$ as a function of temperature in absence of magnetic field and under magnetic fields of 3 T and 9 T. (c) Magnetic entropy change of LaCaScIrO$_6$ in absence of magnetic field as a function of temperature.

**FIG. 8**: (a) μSR time-spectra in LaCaScIrO$_6$ measured in a transverse-field of 3mT on GPS. (b) Temperature dependence of the magnetic volume fraction extracted from the transverse-field μSR. (b) (inset) Internal field-distribution in LaCaScIrO$_6$ collected in a transverse-field of 7800G at temperatures of 300K (purple) and 15K (green). The darker lines are the fits, and the thick brighter lines are the Fourier transforms of the time-spectra.

**FIG. 9**: (a) Zero-field μSR time-spectra in LaCaScIrO$_6$, T =1.6, 9 and 40K in the time-range 0-8 μs. The two component structure at low-temperature is evident, and one sees the high-temperature adopting a more Gaussian shape. (b) Time-range 0-0.3 μs, showing the early time of the spectra in (a) with reduced data binning. The Gaussian Kubo-Toyabe nature of the fast relaxation at low-temperature is evident.

**FIG. 10**: Temperature dependence of zero-field μSR Gaussian and exponential relaxation rates, and the stretch parameter β, in LaCaScIrO$_6$. (a) Gaussian Kubo-Toyabe relaxation rate $\sigma_{KT}$ below T =15 K. (b) Exponential relaxation $\Lambda_f$. The values for T <7.5K are small, but non-zero. (c) Slow-component exponential relaxation rate $\Lambda_s$. (d) The stretch parameter β for the slow relaxation component. Below T =15 K, β was fixed to 1, i.e. purely exponential relaxation for the slow component.

**FIG. 11**: Longitudinal-field measurements of the μSR time spectra in LaCaScIrO$_6$ at T =1.6K. Time-spectra in the time range 0-0.5 μs clearly demonstrate that the Gaussian part of the time-spectra is completely decoupled in fields larger than 2500 G.

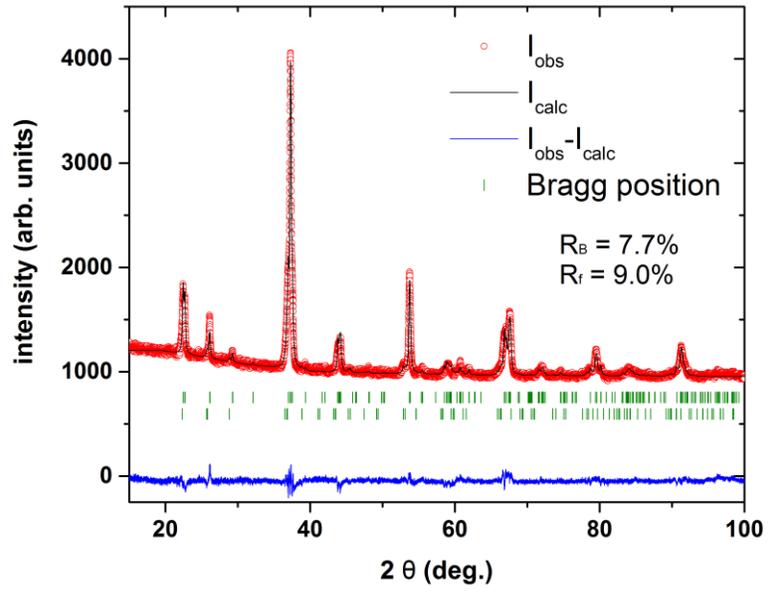

**Fig. 1.**

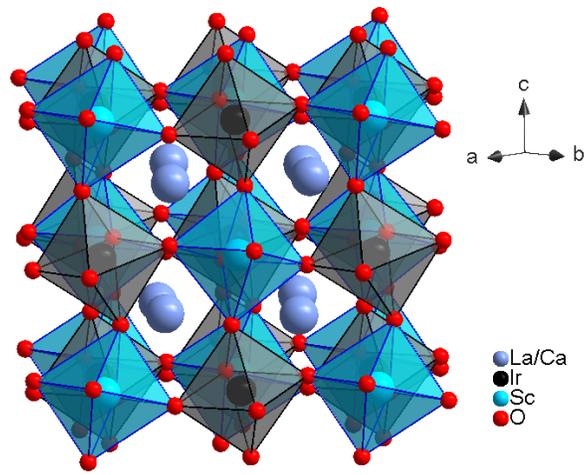

**Fig. 2.**

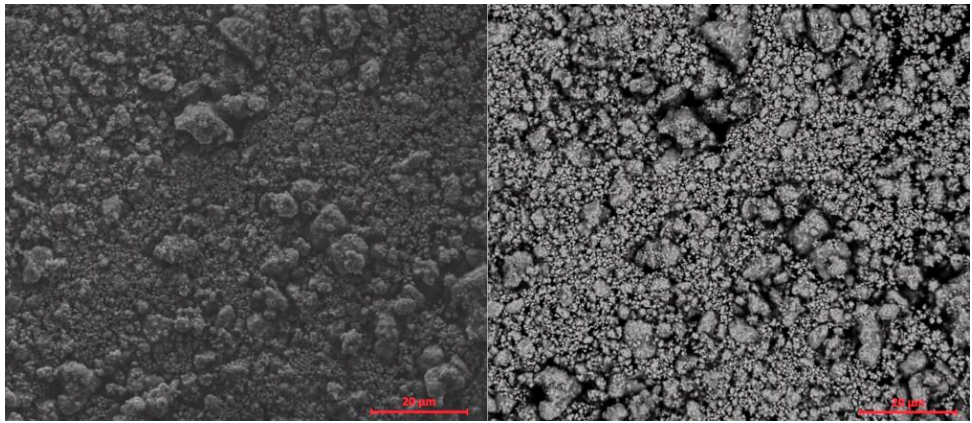

**Fig. 3.**

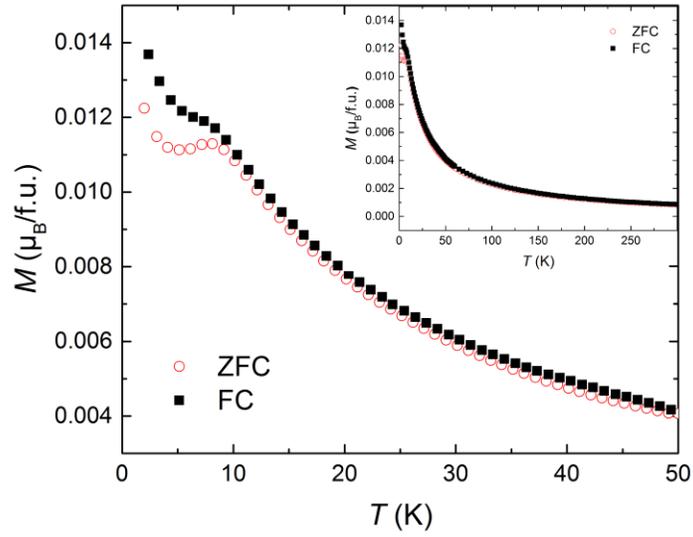

**Fig. 4.**

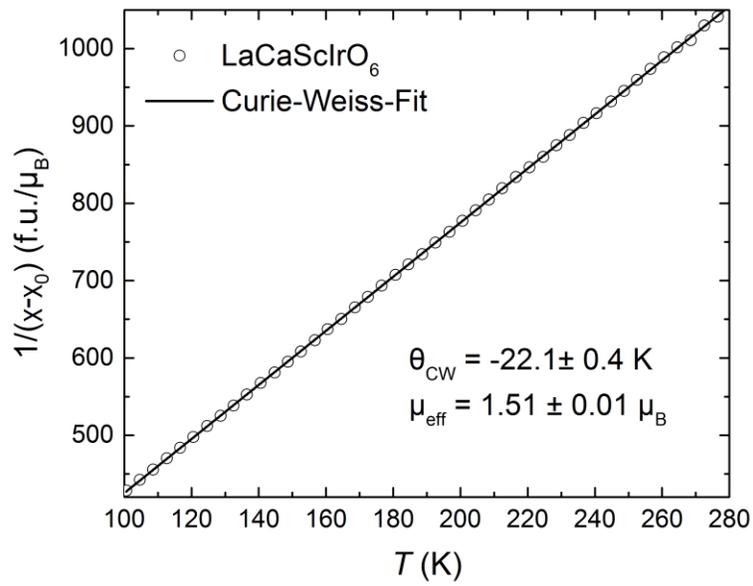

**Fig. 5.**

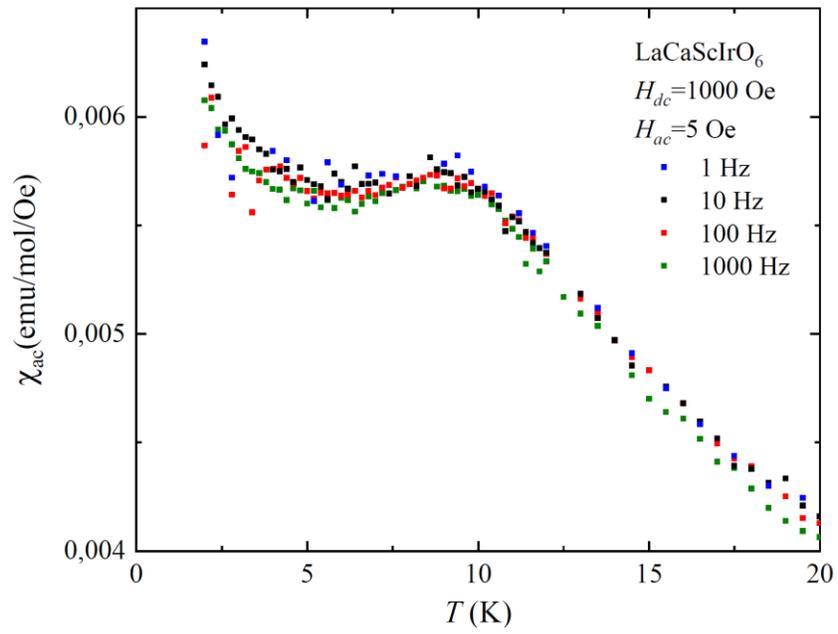

**Fig. 6.**

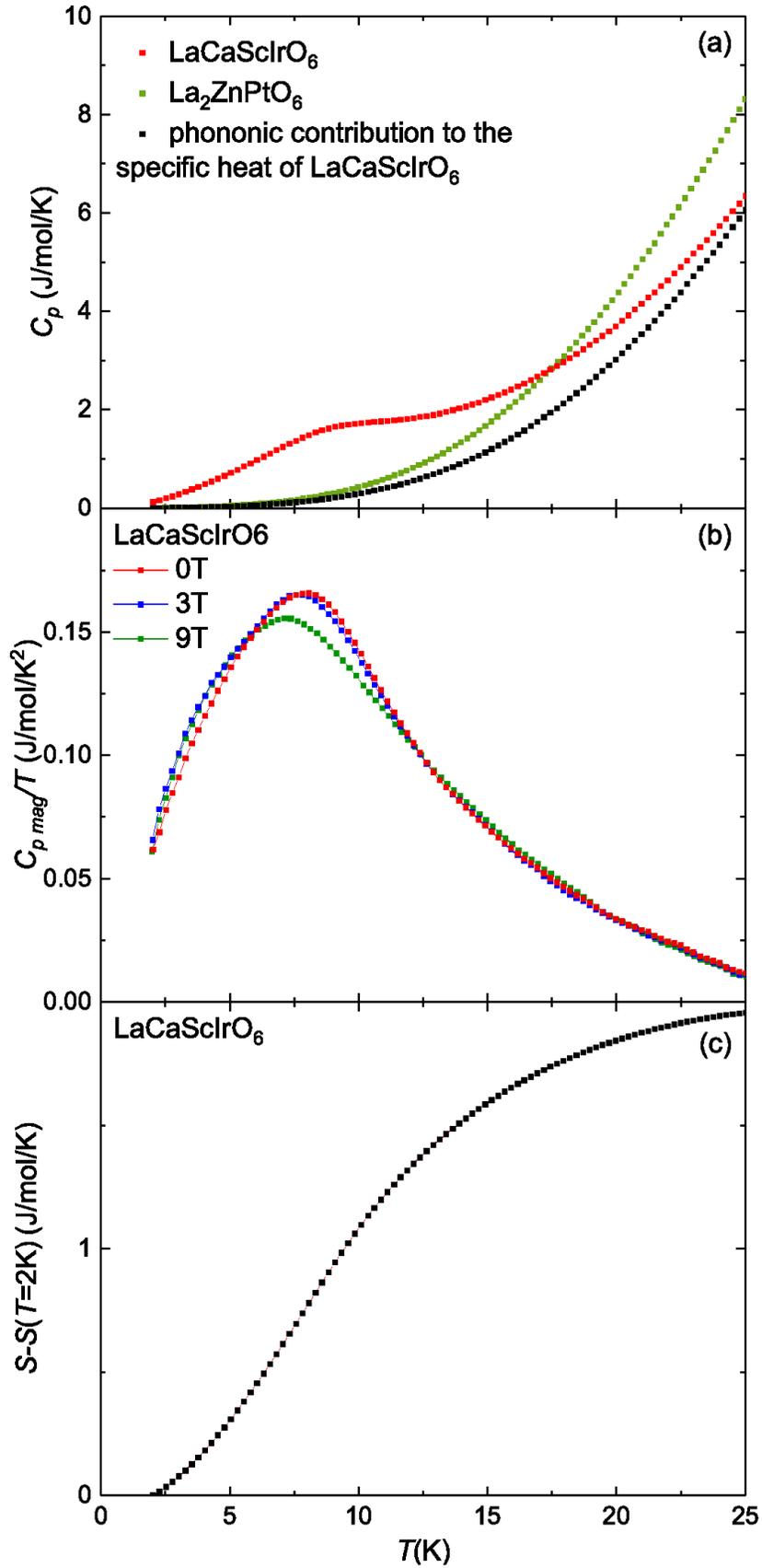

Fig. 7.

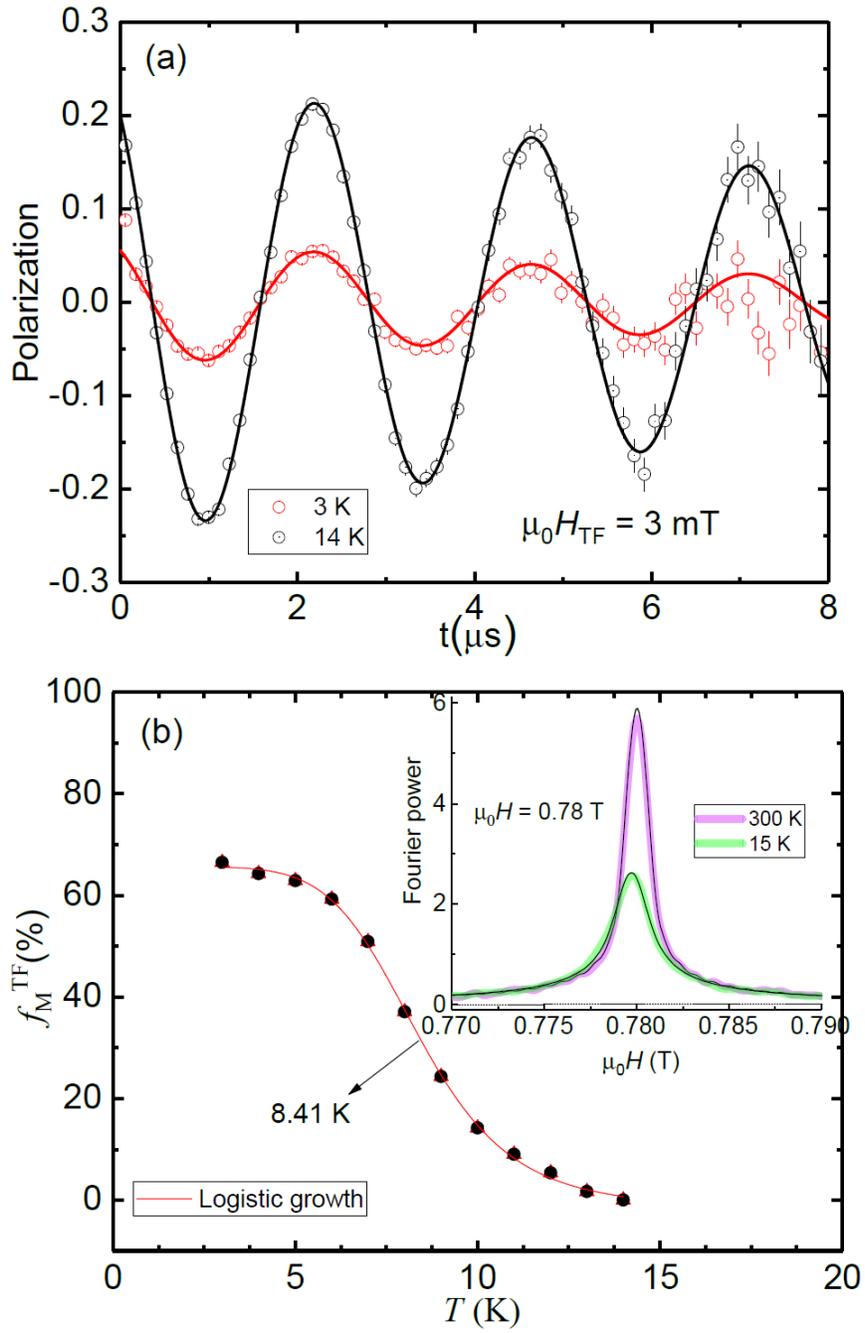

**Fig. 8.**

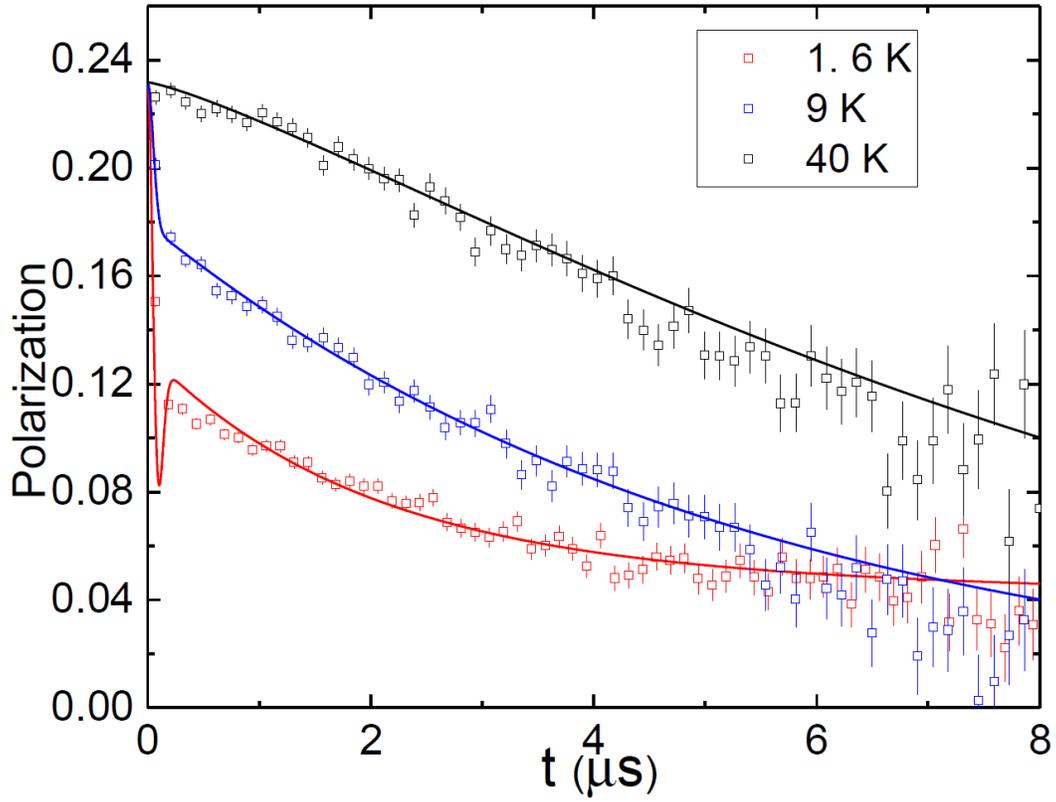
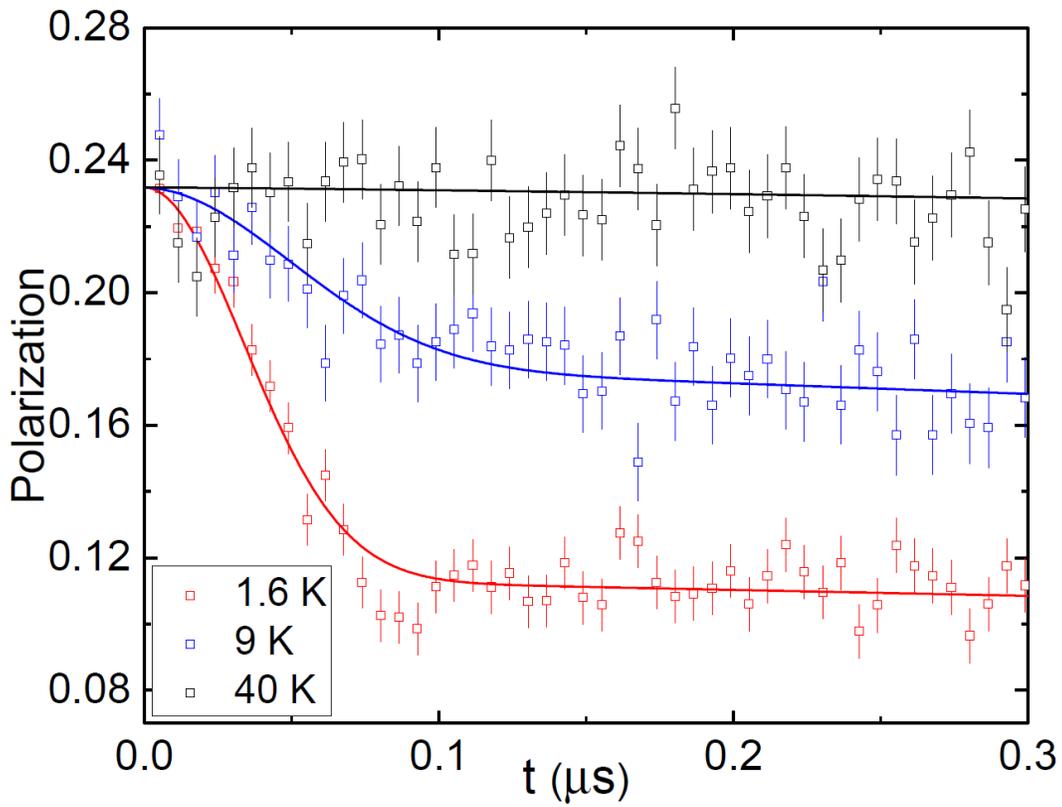

**Fig. 9.**

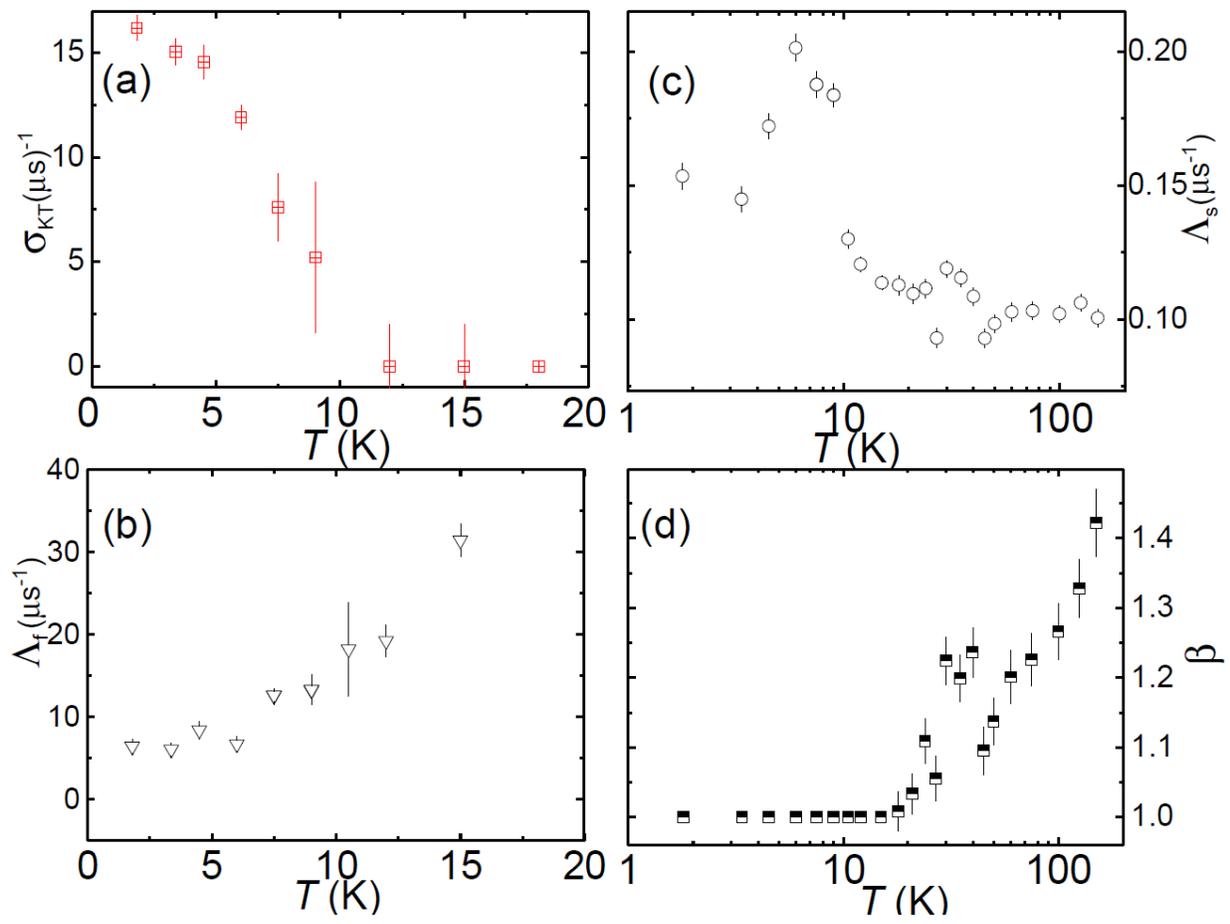

Fig. 10.

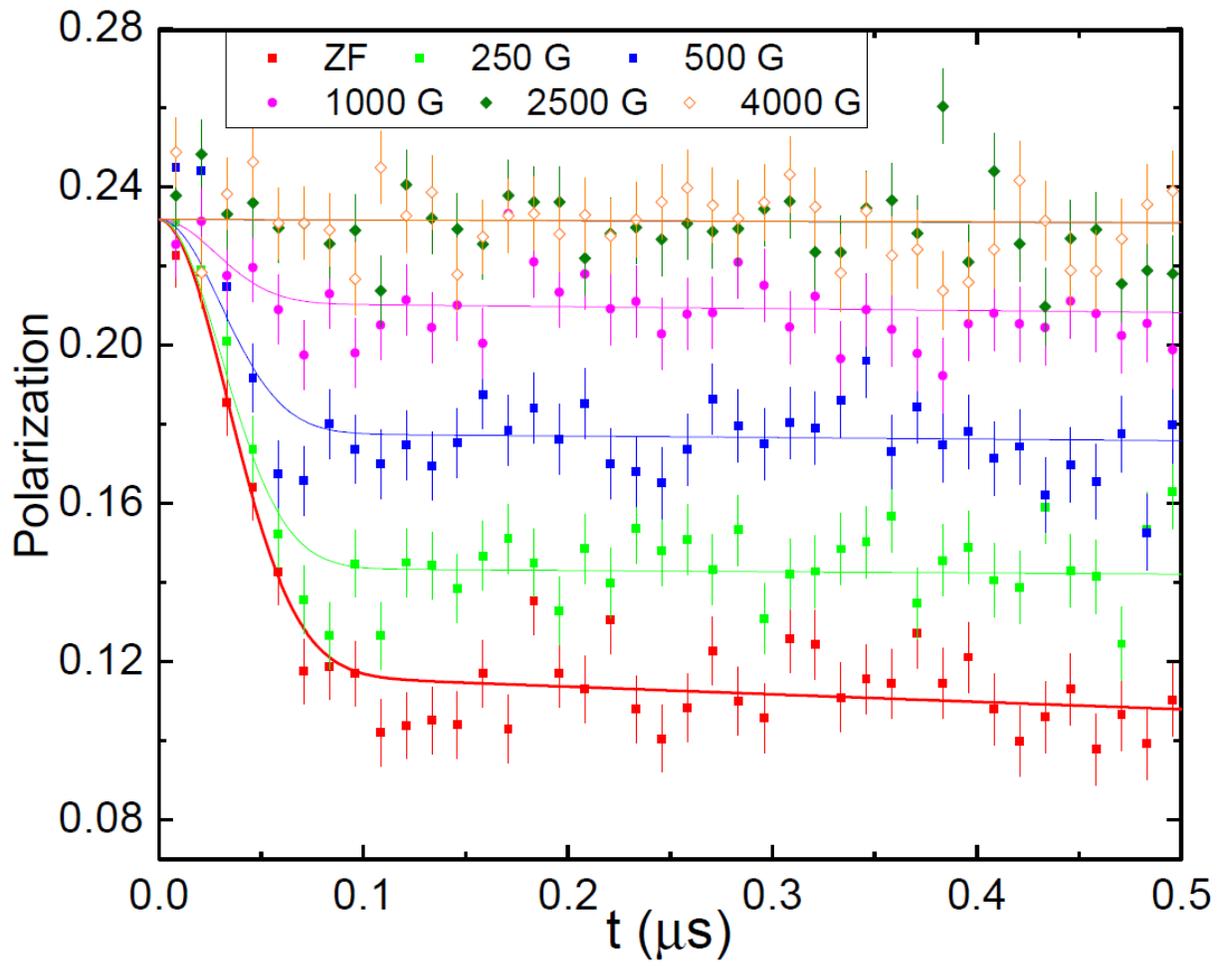

Fig. 11.